# Analysis, Review and Optimization of SONET/SDH Technology for today and future aspects

*Gourav Verma, **Deepika Ramaiya
*Electronics and Communication Engineering Department*
*Northern India Engineering College, Shastri Park, New Delhi.*

*Abstract*-- This paper dedicated to analysis and review of literature for today's technology and future aspects of optical networks. This in depth analysis of today's SONET/SDH Architecture and Reconfigurable structures for SONET rings has been discussed so that one can formulate the next generation SONET/SDH networks. Network layers are analyzed for their design and issues of researches, while dense wavelength division multiplexing equipment has been deployed in networks of major telecommunications carriers for a long time, the efficiency of networking and relation with network control and management have not caught up to those of digital cross-connect systems and packet-switched counterparts in higher layer networks. In this paper, focus on issues by understanding the current structure of the SONET/SDH Layers, its connection to other network technology layers. It will be useful for current OPMA.

*Keywords*--SONET/SDH, STS, Optical carrier, FPGA for SONET, ARM for SONET.

NOMENCLATURE

| | |
|---|---|
| B-DCS | Broadband digital cross-connect system. |
| BoD | Bandwidth on demand. |
| CCAMP | Common control and measurement plane. |
| CMIP | Common management information protocol. |
| CLI | Command line interface. |
| CMISE | Common management information service. |
| CO | Central office. |
| CORBA | Common object request broker architecture. |
| DARPA | Defense Advanced Research Projects Agency. |
| DCS | Digital cross-connect system. |
| DWDM | Dense wavelength division multiplexing. |
| EMS | Element management system. |
| E-NNI | External network-to-network interface. |
| EVC | Ethernet virtual circuit. |
| FEC | Forward error correction. |
| FEC | Forwarding equivalence class (used in MPLS). |
| FXC | Fiber cross connect. |
| Gb/s | Gigabits per second. |
| IETF | Internet Engineering Task Force. |
| GMPLS | Generalized multiprotocol label switching. |
| GUI | Graphical user interface. |
| IOS | Intelligent optical switch. |
| ITU-T | International Telecommunication Union-Telecommunication Standardization Sector. |
| MIB | Management information base. |
| MPLS | Multiprotocol label switching. |
| MPLS-TE | MPLS-traffic engineering. |
| | Muxponder Multiplexer + transponder. |
| NE | Network element. |
| NMS | Network management system. |
| OIF | Optical Internetworking Forum. |
| OMS | Optical mesh service. |
| OSPF | Open shortest path first. |
| OSS | Operations support system. |
| OT | Optical Transponder |
| OTN | Optical transport network. |
| PCE | Path computation element. |
| PMD | Polarization mode dispersion. |
| QPSK | Quadrature phase shift keying. |
| REN | Research and education network. |
| ROADM | Reconfigurable optical add/drop multiplexer. |
| SNMP | Simple network management protocol. |
| SONET | Synchronous Optical NETwork. |
| SRLG | Shared risk link group. |
| TDM | Time division multiplexing. |
| TL1 | Transaction language 1 |

I. INTRODUCTION

Much of the global transport network infrastructure in placetoday is based on the SONET/SDH technology [1],[2]. Thistechnology uses a bandwidth hierarchy indicated by STS-n,where $n = 1, 3, 12, 48...$ The basic unit in this hierarchy isthe STS-I channel, which corresponds to 51.84 Mbps of bandwidth.SONET was originally developed to support voice traffic. The key role of optical networks as the transport infrastructureis to carry client traffic between client networks.Client traffic can be either circuit traffic, e.g., synchronousoptical network (SONET) circuits and asynchronous transfermode (ATM) virtual path/virtual channels, or packet traffic,e.g., Internet protocol (IP) packets, which can be characterizedas traffic flows by forwarding equivalence classes. Optical Network and Management System defining in a SONET/SDH is based on optical layer Sections. The Optical layer is almost the lower level layer for SONET/SDH.

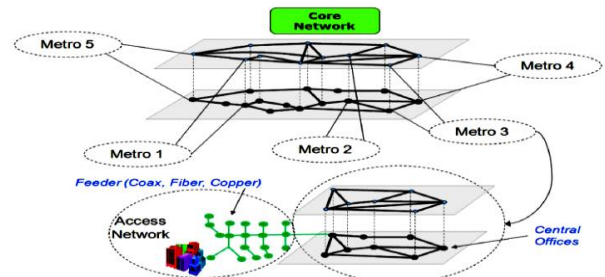

Fig. 1: Terrestrials network layers and segmentations

Network Management and control is addressed in a broad range of bodies such as standard organizations, forums research collaborations, conferences, and journals.

In Section II, the management and control of SONET/SDH layered structure is analyzed, Section III, discusses about the services, SONET can handle like Ethernet, Voice and other important services. The Concept of the paper is to discuss the general layout of the optical layer for defining the future aspects and the problems associated with today's model. Section IV provides us the information about Evolution and structure of today's optical layer. Section V explores current research into evolution of the optical layer, including our assessment of its most likely evolution path.

## II. Management and Control of Optical network

The Management of optical layer as specified in Fig.1. can be seen as layered layout. Networks are organized in two domains nodes and links. The nodes in that case (SONET/SDH) are ADM's Regenerators, Multiplexers and De-multiplexers, ROADM, DCSs etc., about the links we can say about the carriers, OFCs, etc. This can be understood by Network Segments and layers.

### A. Network Segment and Layers

Fig.1. illustrates how we conceptually segment a large national terrestrial network. Large telecommunications carriers are organized into metropolitan (metro) areas and place the majority of their equipment in buildings called COs. Almost all COs today are interconnected by optical fiber. The access segment of the network refers to the portion between a customer location and its first (serving) CO. Networks are further organized into network layers that consist of nodes (switching or cross-connect equipment) and links (logical adjacencies between the equipment), which we can visually depict as network graphs vertically stacked on top of one another. Links (capacity) of a higher layer network are provided as point-to-point demands (also called traffic, connections, or circuits, depending on the layer) in lower layer networks.

### B. Network Layers

Fig. 2. (borrowed from [10]) is a depiction of the core network layers of a large carrier. It consists of two major types of core services: IP (or colloquially, Internet) and private line. Space does not permit us to describe these layers and technologies in detail. We refer the reader to [6] and [14] for background. As one observes, characterizing the traffic and use of the optical layer is not simple because virtually all of its circuits transport links of higher layer networks.

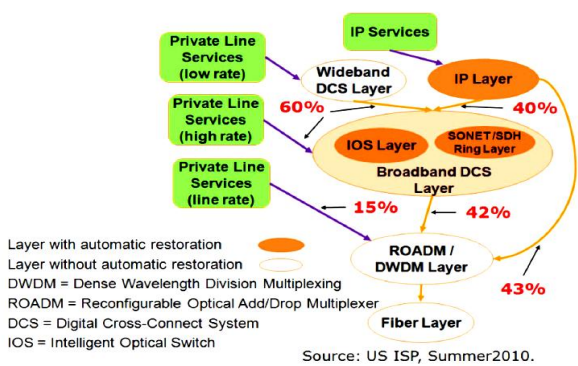

Fig. 2: Core Segment Network layers

As expressed earlier, many industries sweep up the equipment that constitutes the nodes of the upper layer networks of Fig. 2. (such as DCSs) into a broader definition of optical equipment. So, do not attempt to cover network management and control for all these different types of equipment in this paper. Instead, we focus on the definition of optical layer to include legacy point-to-point DWDM systems and newer ROADMs, plus the fiber layer over which they route. One can note that because of the ability to concentrate on today's technology, many vendors enable combinations of these different technology layers into different plug-in slots of the same box(e.g., a DWDM optical transponder on a router platform).

### C. Today's Optical layer

The ITU-T has defined various areas of network management. The area of performance management is also relevant, but applies more to packet networks; therefore, here for simplicity lump relevant aspects of optical performance management into the area of fault management. In the previous section, we discussed provisioning, which is a combination of configuration management and connection management.

#### 1. Legacy DWDM Systems

Clearly, the control plane and network management capabilities of early DWDM systems were simple or non-existent. Virtually all the fault management (alarms) of these systems is based on SONET/SDH protocols from the client signals. Legacy point-to-point DWDM systems were generally installed with simple text-based network management interfaces and a standardized protocol. An example is Bellcore's TL1 [7]. However, for DWDM systems, there is usually an internal communications interface, usually provided over a low rate sideband wavelength (channel). Besides enabling communication between the NEs, this channel is used to communicate with the inline amplifiers. The protocol over the internal communications channel is proprietary.

#### 2. ROADMs

A few EMSs (even sometimes just one) are often used to control the entire vendor sub network, even if the network is scattered over many different geographical regions. Furthermore, the EMS provides an interface to an OSS, typically called a northbound interface using protocols such as CMISE, SNMP [8], CORBA, or XML [9]. Also of interest is that many EMSs use TL1 for their internal protocol with their NEs because it simplifies the implementation of an external TL1 network management interface for those carriers who require it. Firmware or software in the transponders is used to encapsulate client signals of different types(e.g., SONET, SDH, Ethernet, and Fiber Channel) into the internal OTN signal rates.

EMSs can automatically route and cross connect a circuit between a pair of specified transponder ports. Here, the EMS chooses the links and the wavelength, sends cross-connect commands to the individual NEs, monitors status of the circuit request, and reports completion to the northbound interface. The NMS has two main functions: 1) assist planners in the engineering aspects of building or augmenting vendor ROADM sub networks over existing fibers and locations and

2) simulate the paths of circuits over a deployed vendor sub network, taking into account requirements for signal quality.
To summarize as on:
1) The NMS/EMS interaction can be laborious;
2) There may be no flow through from OSS to EMS(via northbound interface);
3) Many portions of the circuit order require manualsteps, such as manual cross connection (patch panel)due to intermediate regeneration or crossingof vendor sub networks;
4) Even with semi-automated or fully automated crossconnection (which is an order of magnitude fasterthan above), optical signal settling times can belong compared to cross-connect speeds in higherlayer networks.

Finally, fault management is similar to that of the point-to-point DWDM system, except that all newerROADM internally use OTN encapsulation of the circuitsand, as a result, the alarms identify affected slots and portsin terms of the OTN termination-point information models and alarm specifications.

*3. Integrated Interlayer Network Management*
We revisit two of the key network characteristics highlighted in the introduction, namely network layeringand restoration. Because today restoration is typicallyperformed at higher layer networks, outages that originateat lower layers are more difficult to diagnose and respond.For example, an outage or performance degradation of a DWDM amplifier or a fiber cut can sometimes affect ten ormore links in the IP layer, while the failure of an intermediatetransponder may affect only one IP-layer link andbe hard to differentiate from outage of an individual router port. IP backbones have traditionally relied on IP-layer re-convergence mechanisms, (generally called internalgateway protocols), such as OSPF [18] or more explicitrestoration protocols such as MPLS fast reroute andMPLS-TE [19].

With IP routing protocols that do not take into account linkcapacity (e.g., OSPFVbut note a capacity-sensitive versioncalled OSPF-TE has been defined), losing a significantnumber of component links of a link bundle (but not all),would normally result in the normal traffic load on thislink being carried on the remaining capacity, potentiallyleading to significant congestion. In recent years, router technologies have been adapted to handle such scenarios, shuttingdown the remaining capacity in the event that the linkcapacity drops below a certain threshold. Routers will detect outages which occur anywhere on alink, be it due to a port outage of the router at the remoteend of the link, an optical amplifier failure, or fiber cut.

However, the IP and optical layersare typically managed by very distinct work groups or evenvia an external carrier (e.g., leased private line). In theevent of an optical-layer outage, the alarm notificationswould also be created to the optical maintenance workgroups. Thus, without sophisticated alarm correlation mechanisms between the events from the two different layers, there can be significant duplication of trouble shooting activities across the two work groups. Efficient correlation of alarms generated by the two different layers can ensure that both work groups are rapidly informed ofthe issue, but that only the optical-layer group neednecessarily respond as they would need to activate the necessary repair.

3. *Metro Segment*
In contrast to the core segment, metro networks have considerably smaller geographical diameter. A circuit path can involve complex access provisioningon distribution/feeder fiber followed by long sequencesof patch panel cross connects in COs. For example, if a circuit requires 15 manual cross connectsover direct fibers and only one section of automated crossconnection over ROADMs, it is hard to prove the businesscase for the ROADM segment since overall cost is nothighly impacted. Length constraints prevent us fromdelving into more detailed metro issues.

### III. ETHERNET OVER SONET/SDH

*A. Ethernet*
Ethernet is a connectionless packet-switching technology, defined by a set of physical and data link specifications, functions and protocols originally developed for computernetworking. In 1985, the 802.3standardization committee of the *Institute of Electrical andElectronics Engineers* (IEEE) published its Ethernet standardwith the title *IEEE 802.3 Carrier Sense Multiple Access withCollision Detection (CSMA/CD) Access Method and PhysicalLayer Specifications* [11].Ethernet is the dominant technology in computer*Local Area Networks* (LANs) [13], [15], [16]. Ethernet standardIEEE 10BASE-T [20] provided up to 10 Mbit/s in one unshieldedtwisted pair using baseband Manchester line coding [17]. Themaximum segment size is 100 meters. The 10BASE-Tstandard became widely adopted to transport *Internet Protocol* (IP) [26x] datagrams, which are accommodated on Ethernetframes.

An Ethernet frame contains [11]: a 7-octet *Preamble*, which is a sequence of alternated 0s and 1s used to establish bitsynchronization between source and destination hardware; a 1-octet *Start-of-Frame-Delimiter* (SFD), which indicates the firstbit of the rest of the frame; 12 octets of *Source* and *Destination*.

*Media Access Control* (MAC) data link sublayer addresses; a 2-octet *Length/Type* that takes one of two meanings: to indicateframe length in IEEE 802.3 standards (which is limited to1518 octets), or to indicate which network layer protocol isbeing carried in the frame, in order to maintain compatibilitywith the DIX standard; 46 to 1500 octets of MAC client dataand/or padding; and 4 octets of *Frame Check Sequence* (FCS) which is a 32-bit *Cyclic Redundancy Check* (CRC). For theCSMA/CD protocol to function correctly, a minimum MACframe size is required, and thus padding can be added to theframe if needed. Also, IEEE 802.3 [11] defines an *Inter-Packet Gap* (IGP) between Ethernet frames to provide adequate recovery timesfor procedures in the link and physical layers, such as cycling circuitry from transmit to receive mode in half-duplex operation.

The IGP for 10BASE-T standard is 9.6 μ seconds, while it is 0.96μs for 100BASE-T. This is equivalent to 12 bytes of mission time in these standards. The IGP is related to the *Inter-Frame Spacing* (IFS). According to [21], the IFS is the sum of at least 12 bytes of IGP, plus a 7-octet *Preamble* and a 1-octet SFD. Also, Ramamurti et al. [22] discusses IFS and IGP, and IGP use for rate adaptation in EoS. *Gigabit Ethernet* (GbE) was developed to interconnect 10/100 Mbit/s switches and to provide higher data rates. The goal of 10GbE was to cover distances from 300 meters to 40 km. Only optical physical layer options were defined. In addition, 10GbE does not support half- duplex operation or CSMA/CD; all operation is in full-duplex mode.

B. *EoS*

EoS stands for Ethernet over SONET. That is framing the Ethernet frames over SONET frames. As shown in Figure 3, there are several ways by which IP data can be supported over SONET. The first approach is to use IP over-ATM-over-SONET using *AALS* (ATM Adaptation Layer *5*) [27]. Under POS, PPP-encapsulated IP packets are framed using the High-Level Data Link Control (HDLC) protocol and are mapped into SONET. The basic function of HDLC is to provide framing, i.e., delineation of the PPP-encapsulated IP packets across the synchronous transport link.

Another method for transporting IP data over SONET is to use the Generic Framing Procedure (GFP), which encapsulates Ethernet frames and then map them into SONET frames [29]. It is currently considered the most popular framing procedure for supporting Ethernet-over-SONET (EoS), being required to support EoS and virtual concatenation.

EoS has been gaining popularity in point-to-point and multi-point LAN interconnections [30]. EoS with *virtual concatenation* [1] utilizes the existing SONET infrastructure with only the edge nodes (source and destination). It also facilitates dynamic link upgrade without additional hardware using the Link Capacity Adjustment Scheme (LCAS) [31]. In VCAT, many STS-n channels, belonging to possibly different Optical Carriers (OCs)', can be concatenated between the source and destination to support Ethernet connectivity. These STS-n channels form a Virtually Concatenated Group (VCG). A key factor that impacts the dynamic establishment of a new STS-n channel is the differential delay between all existing STS-n channels of the VCG and the newly added STS-n channel.

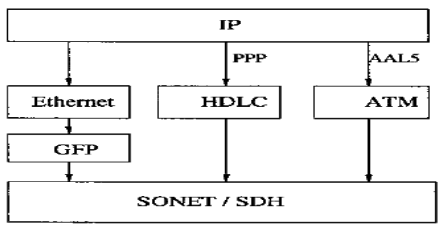

Fig. 3. Methods for transporting IP over SONET

This bound is determined by the amount of high-speed memory available at the edge node that stores the incoming SONET frames from different OCs.

The smallest SONET payload slot that can carry such traffic is STS-48(2.5 Gbps), which results in bandwidth wastage of about 60%. A solution to avoid this problem is the concatenation or concatenated payloads. Two methods for concatenation are available are [1]: Contiguous and Virtual Concatenation. Both methods provide aggregate bandwidth of X times the bandwidth of the STS-n channels at the termination (n = 1, 3, 12, 48...). Contiguous concatenation maintains contiguous bandwidth throughout the transport path, and constituent STS-n channels of the concatenated payload cannot be individually and independently routed as shown in fig. 4.

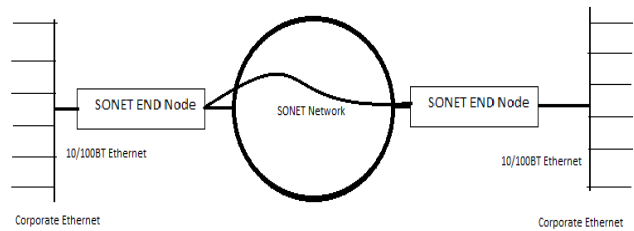

Fig. 4. EoS setup with Contiguous Concatenation

In contrast, virtual concatenation splits the aggregate bandwidth into several VCs that are independently established between the two end points, as shown in fig. 5. The routes of these VCs may or may not overlap. Whereas, contiguous concatenation requires concatenation functionality at each network element, VC requires such functionality only at the path termination equipment.

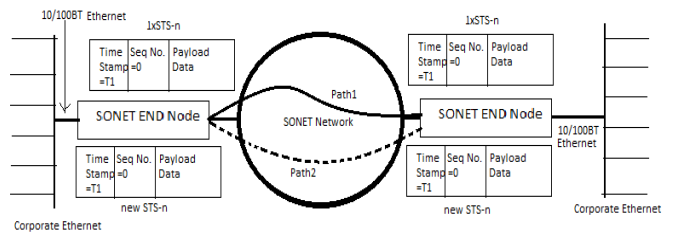

Fig. 5. EoS setup with Virtual Concatenation

C. *Support of ATM, POS and GFP*

Compared to carrying plesiochronous traffic, carrying asynchronous transfer mode [17] (ATM), packet over SONET [18] (POS), or generic framing procedure [19] (GFP) traffic is a piece of cake. For an STS-1, the payload consists of 9 rows and 87 columns, one of which is the POH, and two are fixed stuff (columns 30 and 50 numbered from the POH). This leaves 84 columns by nine rows for payload.

Beyond that one requirement, the payload of the SONET frame is simply viewed as an octet transport mechanism. As an example, ATM cells are taken one octet at a time with each octet placed in the next available octet in the SPE without regard for any boundaries in the cell or the SPE, other than maintaining octet alignment. POS and GFP are handled in exactly the same way.

As an aside, note that the SPE of an STS-1 always has columns 30 and 59 of the SPE stuffed and unavailable for payload traffic. If a customer had the option of putting traffic into three STS-1s or one STS-3c, it would be better to choose the STS-3c. Let's see why. The SPE of an STS-3c consists of 261 columns (270 columns minus 9 columns for transport overhead). The POH will take one column of the SPE leaving 260 columns for user traffic. If the customer used three STS-1, he/she would receive three times 84 columns of payload, or only 252 columns compared to 260 columns for the STS-3c.

Eight columns of payload is equal to a little more than 4.6 Mbps, or the equivalent of about three DS-1s. It's one of the oddities of SONET/SDH that part of this extra bandwidth is only available at STS-3c and not at higher levels of SONET/SDH.

IV. STRUCTURE OF TODAY'S OPTICAL LAYER

A. *Reconfigurable Optical Add/Drop Multiplexer (ROADM)*

Today legacy point-to-point DWDM systems still carry older circuits and sometimes are used for segments of new circuit orders, especially lower rate circuits. However, most large carriers now augment their optical layer with ROADMs. In contrast to a point-to-point DWDM system, a ROADM can interface multiple fiber directions (or degrees). This has encouraged the development of more flexibly tuned transponders (called nondirectional or steerable) and the ability to perform a remotely controlled optical cross connect (e.g., through wavelength-selective cross connects). See [14] and [41].

A ROADM can optically (i.e., without electrical conversion) cross connect the constituent signals from two different fiber directions without fully demultiplexing the aggregate signal (assuming they have the same wavelength). This is called a transit or through cross connection. Or, it can cross connect a constituent signal from a fiber direction to an end transponder, called add/drop cross connection. All ROADM vendors provide a CLI for communication with a ROADM and an EMS that enables communication with a group of ROADMs. These network management and control systems are used to allow personnel to perform optical cross connects. Possibly the same personnel perform this request by manually fibering jumpers between the appropriate ports on the patch panel itself. See [14]. If an FXC is deployed, then the installation personnel must still fiber the transponder ports and client equipment to the FXC, but when the provisioning order is given, the FXC can cross connect its ports under remote control. However, today, there are few FXCs deployed in large carriers; therefore, in this section, one will assume the patch panel dominates, but return to the FXC in our last section.

Here the list four broad categories of provisioning steps in the core segment. In many cases, a circuit order may require steps from all four categories.

1) Manual: installation personnel visit CO, install cards and plug-ins, and fiber them to the patch panel.
2) Manual: installation personnel visit CO and cross connect ports via the patch panel.
3) Semiautomatic: Provisioned request optical cross connects via a CLI or EMS.
4) Fully automated: an OSS is fed a circuit path from a network planner or planning tool and then automatically sends optical cross-connect commands to the CLI or EMS.

Carriers are mostly doing category 3) today. Fig. 6 depicts a realistic example within the optical layer of Fig. 2, where a 10-Gb/s circuit is provisioned between ROADMs A-G. For example, this circuit might transport a higher layer link between two routers which generate the client signals at ROADMs A and G.

There are two vendor sub networks in this example, where a vendor sub network is defined to be the topology of vendor ROADMs (nodes) from a given equipment vendor plus their inter connecting links (fibers). Because DWDM systems from different vendors do not generally support a handoff (interface) between light paths, for a circuit to cross vendor sub networks requires. add/dropping through transponders. The ROADMs in this example support 40-Gb/s channels/wavelengths. Another complicating factor in today's networks is the evolution of the top signal rate over the years. In this example, need to multiplex the 10-Gb/s circuit into the 40-Gb/s wavelengths.

DWDM equipment vendors provide a combo card, colloquially dubbed a mux ponder, which provides both TDM (dubbed "mux") and transponder functionality. To provision our example 10-Gb/s circuit, must first provision two 40-Gb/s channelized circuits (i.e., they provide 4-10b/s sub channels), one in each sub network (A-C and D-G). Furthermore, because of optical reach limitations, the 40-Gb/s circuit must de-multiplex at F and thus traverse two light paths in the second sub network.

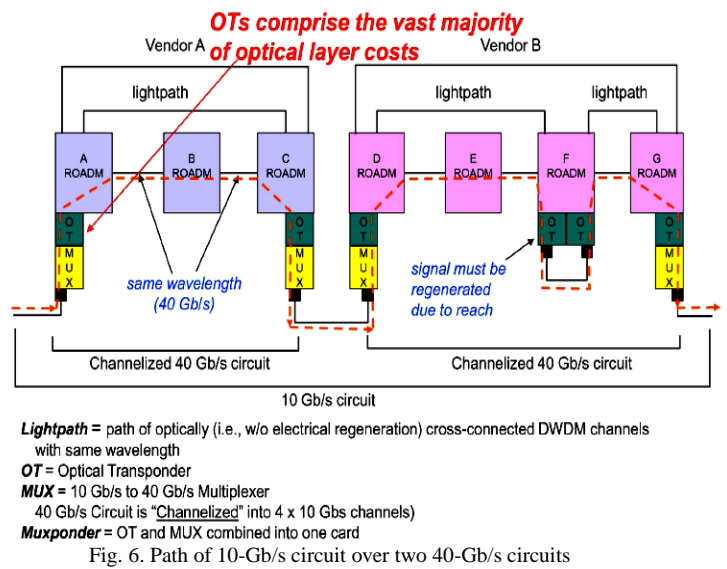

Fig. 6. Path of 10-Gb/s circuit over two 40-Gb/s circuits

An interesting observation from Fig. 6 is thatbecause of the logical links created at each layer; sometimeslinks at a given layer appear to be diversely routed, when infact they converge over segments of lower layer networks.

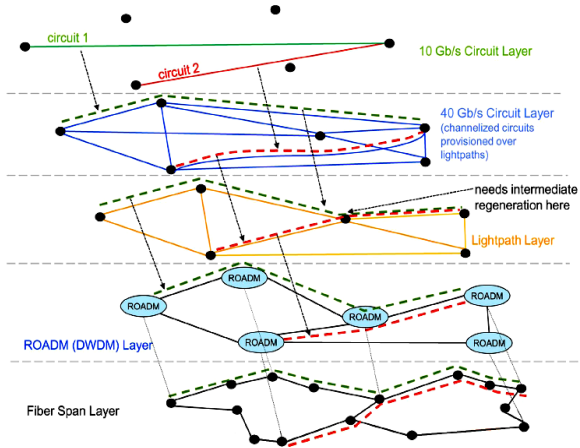

Fig. 7: Layers of SONET

## V. NEXT GENERATION SONET/SDH SCHEMES

Powered with the knowledge of the current status of SONET/SDH, EoS and Packet over the SONET In this paper, one can define or analyze a domain of next generation SONET schemes. Take SONET OC-768 as anexample, 40 Gb/s digital signal transmission (50 Gb/s when using some suggested forward error correction schemes) will require a package that works fairly well up to 60 GHz—the third order harmonic of the equivalent primary frequency. Traditionally package designs above 20 GHz have been focused on narrowband applications, and packaging design options for wideband performance become very limited at higher frequency range, especially for off-chip interconnections.

*A. Network Control and Management Gap*

We summarize the following observations about theoptical layer in today's carrier environment.1) The optical layer can require many manual stepsto provision a circuit, such as NMS/EMS circuitdesign coordination, crossing vendor subnetworks,and intermediate regeneration because ofoptical reach limitations.2) Even the fully automated portions of provisioning an optical-layer circuit are significantly slowerthan its higher layer counterparts.3) Evolution of the optical layer has been heavilymotivated to reduce costs for interfaces to upperlayer switches. This has resulted in a simple focus to increase B-rate and reach.4) Restoration is provided via higher network layersand, thus, planning, network management, andrestoration must work in a more integratedfashion across the layers.5) No large-scaled dynamic services have been implementedthat would require rapid connectionmanagement in the optical layer.

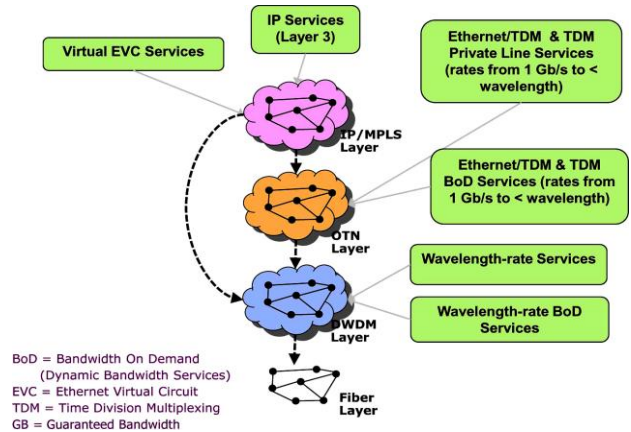

Fig. 8.Future scope of optical layer

*B.Technology Evolution of the Optical Layer*

Optical and WDM transport technology has undergoneimpressive technological advancement in the past 15 years.As previously described, DWDM technology started with afew wavelengths, low bit rates, and limited point-to-point networking. Today, ROADM systems are being deployedwith rates of 100 Gb/s, 80 wavelengths, and lightpathswith1000–1500-km reach. Thus, the principal drivers for higher "rate"wavelengths will not be as intense as in the past. Thetop-rate interface on packet switches has steadily evolvedin steps, e.g., 155 Mb/s, 622 Mb/s, 2.5 Gb/s, 10 Gb/s,40 Gb/s, and 100 Gb/s. DWDM channel rates have matched.

*C. Methods for Fully Automated Provisioning*
See [22] and [14] for optimization algorithms forsizing and placing pools of transponders. Both of theseconcepts are key components of the CORONET project [36]. The main purpose of the NMS is to theoreticallyroute (also called "design") a circuit over a path of light paths (including selection of spare wavelengths) and intermediatetransponders (if needed) to ensure that adequatespare channel capacity exists and that signal quality isprovided. The authors and collaborators have derived andimplemented a process in AT&T's network to automate theNMS portion of the provisioning step.

*D. Business Case for Optical-Layer Evolution*
After over a decade of technical development, while optical-layer capacity, connectivity, cost improvements, and signal quality have enjoyed great advancement, optical management and control has evolved more slowly. We have shown this is clearly not due to lack of R&D, both in advanced network architectures and protocols [39]. The authors feel that mostof these advances will eventually be implemented becauseof 1) the leveling of core IP traffic growth (and thus thelack of historically frenzied need for wavelength rate increase);2) continued decline in transponder costs andprices; and 3) advancements in DWDM technologies.However, the key variable will be the rate of this implementation,which will hinge on the ability to prove the business cases.

## VI. Conclusions

From our most of the discussion and analysis of SONET system we have analyzed the current structure of SONET/SDH design issues, the layers, the devices advancements in routing and topology structures. Not only this we discussed about the Ethernet quality improvement and mapping of the Ethernet frame over SONET frame. The differential delays in a routing path are discussed which is one of the key point for distribution of packet timings. Moreover we can say that we have completely analyzed the current structures of SONET/SDH frame works and layout.

With the power of current technology knowledge, now we are able to define the next generation implementation which we have discussed and optimized in section V. This knowledge of the implementation defines the new algorithms and new devices like Reconfigurable Add Drop Mux, DCSs Routing algorithm at highest optimized level.

## References

[1] ITU-T Standard G.707, "'Network node interface for the synchronousdigital hierarchy," ZW0. pp. 118-126.

[2] ANSI Tl.105-2001, "Synchronous optical network (SONET). Basic description including multiplexing smcme, rates md formats," .

[3] E. Rosen, A. Viswanathan, and R. Callon.(2001, Jan.)Multiprotocol label switching architecture. IETF, RFC 3031.

[4] R. Doverspike & P. Magill, "Optical Fiber Telecommunications", in CON, Overlay Networksand Services. Amsterdam, The Netherlands: Elsevier, 2008.

[5] R. Doverspike, K. K. Rama-krishnan and C. Chase, "Guide to Reliable Internet Services and Applications", in Structural Overview of commercial Long Distance IP Networks, C. Kalmanek, S. Misra, and R. Yang, Eds. 1st ed.New York: Springer-Verlag, 2010.

[6] R. Doverspike and P. Magill, "Chapter 13in Optical Fiber Telecommunications V inCommercial Optical Networks, Overlay Networksand Services". Amsterdam, The Netherlands:Elsevier, 2008.

[7] Bellcore, Operations ApplicationMessagesLanguage for Operations Application Messages, TR-NWT-000831.[Online]. Available: http://telecominfo-telcordia.com/.

[8] U. Black, "Network ManagementStandards SNMP, CMIP, TMN, MIBs, and Object Libraries", A. Bittner, Ed.New York: McGraw-Hill, 1995, ISBN: 007005570X.

[9] J. Zuidweg,"Next Generation IntelligentNetworks". Norwood, MA: Artech House,2002.

[10] A. Gerber and R. Doverspike, "Traffic typesand growth in backbone networks, inProc. Conf. Opt. Fiber Communication ", Nat. FiberOpt. Eng. Conf., Los Angeles, CA, Mar. 2011,pp. 1–3.

[11] IEEE, "Carrier Sense Multiple Access with Collision Detection(CSMA/CD) Access Method and Physical Layer Specifications," IEEEStd 802.3-2008, December 26, 2008.

[12] J. Yates and Z. Ge, "Chapter 12 in Guide toReliable Internet Services and Applications,in Network Management: Fault Management,Performance Management, and PlannedMaintenance", C. Kalmanek, S. Misra, and R. Yang, Eds., 1st ed. New York:Springer-Verlag, 2010.

[13] C. Spurgeon, "Ethernet, The Definitive Guide," O´Reilly, ISBN 1-56-592660-9, 2000.

[14] M. D. Feuer, D. C. Kilper, andS. L. Woodward, "Chapter 8 in Optical Fiber Telecommunications V B,inROADMs and Their System Applications Amsterdam," The Netherlands: Elsevier,2008.

[15] M. Ali, G. Chiruvolu and A. Ge, "Traffic Engineering in Metro Ethernet,"*IEEE Network*, pp. 10-17, Mar./Apr. 2005.

[16] W. Stallings, "Data and Computer Communications," Prentice Hall, 7th edition, ISBN 0-13-1000681-9, 2004.

[17] W. Stallings, "Data and Computer Communications," Prentice Hall, 7[th]edition, ISBN 0-13-1000681-9, 2004.

[18] IETF RFC 2328, OSPF Version 2, Apr. 1998.[Online]. Available: http://www.ietf.org/rfc/rfc2328.txt.

[19] IETF RFC 4090, "Fast Reroute Extensionsto RSVP-TE for LSP Tunnel", May 2005.[Online]. Available: http://www.ietf.org/rfc/rfc4090.txt.

[20] X. He, M. Zhu, and Q. Chu, "Transporting Metro Ethernet Servicesover Metropolitan Area Networks", in *Proceedings of the IEEE InternationalConference on Sensor Networks, Ubiquitous, and TrustworthyComputing*, 2006, pp. 178-185.

[21] X. He, M. Zhu, and Q. Chu, "Transporting Metro Ethernet Servicesover Metropolitan Area Networks", in *Proceedings of the IEEE InternationalConference on Sensor Networks, Ubiquitous, and TrustworthyComputing*, 2006, pp. 178-185.

[22] V. Ramamurti, J. Siwko, G. Young, and M. Pepe, "Initial Implementationsof Point-to-Point Ethernet over SONET/SDH Transport," *IEEECommunications Magazine*, pp. 64-70, Mar. 2004.

[23] IEEE, "Type 1000BASE-X MAC Parameters, Physical Layer, Repeater,and Management Parameters for 1000 Mb/s Operation," IEEE 802.3z-1998, 1998.

[24] IEEE, "Media Access Control (MAC) Parameters, Physical Layers, and Management Parameters for 10 Gb/s Operation," IEEE Std 802.3ae-2002, 2002.

[25] IEEE, "Amendment 1: Physical Layer and Management Parameters for 10 Gb/s Operation," Type 10GBASE-T, IEEE Std 802.3an-2006, 2006.

[26] DARPA, "Internet Protocol," RFC 791, Sept. 1981.

[27] **IETF** RFC **1577,** "Classical IP and ARP **over** ATM:' **lune 1994.**

[28] IETF WC **261.5,** "PPP over SONETISDH:' **June 1999.**

[29] ITU-T Standard **G.7041.** "Generic framing procedure:' Feb **2003.**

[30] V. Rammuni, I. Siwko, G. **Young,** and M. Pepe, "Initial implementationsof point-to-point ethernet**ova** SONETISDH transpon:' in *IEEECommunicotionr Magazine.* **2W4, pp.** 64-70.

[31] ITU-T Standard **G.7042.** "Link capacity adjustment scheme for virmally concatenated signals:' 2001.

[32] D. Cavendish, "Evolution of optical transport technologies: from SONET/SDH to WDM:' in *IEEE Communicotionr Magazine,* 2000,pp. 164-172.

[33] P. Ashwood-Smith, Y. Fan, A. Banerjee,J. Drake, J. Lang, L. Berger, G. Bernstein,K. Kompella, E. Mannie, B. Rajagopalan,D. Saha, Z. Tang, Y. Rekhter, andV. Sharma, Generalized MPLSVSignalingFunctional Description, IETF Internet draft,Jun. 2000.

[34] C. Fludger, T. Duthel, D. van den Borne,C. Schulien, E.-D.Schmidt, T. Wuth,J. Geyer, E. De Man, G.-D.Khoe, andH. de Waardt, BCoherent equalizationand POLMUX-RZ-DQPSK for robust100-GE transmission,[ J. Lightw. Technology.,vol. 26, no. 1, pp. 64–72, Jan. 2008.

[35] M. G. Taylor, BCoherent detection methodusing DSP for Demodulation of signaland subsequent equalization of propagationimpairments,[ IEEE Photon. Technol. Lett.,vol. 16, no. 2, pp. 674–676, Feb. 2004.

[36] J. Strand, BIntegrated route selection,transponder placement, wavelengthassignment, and restoration in an advancedROADM architecture,[ J. Lightw. Res.,May 2011.

[37] X. Zhang, M. Birk, A. Chiu, R. Doverspike,M. D. Feuer, P. Magill, E. Mavrogiorgis,J. Pastor, S. L. Woodward, and J. Yates,"Bridge-and-roll demonstration in GRIPhone(globally reconfigurable intelligent photonic network)",[ in Proc. Conf. Opt. Fiber Commun./Nat. Fiber Opt. Eng. Conf., San Diego, CA,Mar. 2010, pp. 1–3.

[38] G. Li, A. Chiu, R. Doverspike, D. Xu, andD. Wang, "Efficient routing in heterogeneouscore DWDM networks,[ in Proc. Conf. Opt.Fiber Commun.Nat. Fiber" Opt. Eng. Conf.,San Diego, CA, Mar. 2010, pp. 1–3.

[39] Optical Internetworking Forum (OIF),External Network-Network Interface(E-NNI) OSPF-Based Routing-1.0(Intra-Carrier) Implementation Agreement,OIF-ENNI-OSPF-01.0, Jan. 17, 2007. [Online].

[40] I. Kaminow and T. Koch, Eds., Optical Fiber Telecommunications III. New York: Academic, Apr. 14, 1997.

[41] P. Palacharla, X. Wang, I. Kim, D. Bihon,M. Feuer, and S. Woodward, "Blockingperformance in dynamic optical networksbased on colorless, non-directionalROADMs", in Proc. Conf. Opt. Fiber Commun./Nat. Fiber Opt. Eng. Conf., Los Angeles, CA,Mar. 2011, pp. 1–3.

[42] Robert D. Doverspike, Jennifer Yates,"Optical Network and management", Proceedings of the IEEE | Vol. 100,No. 5, May 2012, pp.: 1092-1104

[43] Antˆonio Marcos Alberti&RoulienFernandes, "Ethernet-over-SDH: Technologies Review and Performance Evaluation", REVISTA TELECOMUNICAÇÕES, VOL. 13, Nº. 01, MAIO DE 2011, pp: 1-23

[44] SatyajeetS. AhujatTurgayKorkmaztManvanKrunzt, "Minimizing the Differential Delay for Virtually Concatenated Ethernet Over SONET Systems", IEEE 2004.

[45] Lei Shan, MounirMeghelli, Joong-Ho Kim, Jean M. Trewhella, and Modest M. Oprysko," Simulation and Design Methodology for a 50-Gb/s Multiplexer/Demultiplexer Package", IEEE TRANSACTIONS ON ADVANCED PACKAGING, VOL. 25, NO. 2, MAY 2012

[46] ChunshengXin, ChunmingQiao, Sudhir Dixit, "Traffic Grooming in Mesh WDM Optical Networks—Performance Analysis", IEEE JOURNAL ON SELECTED AREAS IN COMMUNICATIONS, VOL. 22, NO. 9, NOVEMBER 2004.

[47] Tope R. Karem and H. Anthony Chan", A Low-Cost Design of Next Generation SONET/SDH Networks with Multiple Constraints.", IEEE 2007.

[48] Ullas Kumar," Synchronizing network elements on SONET/SDH rings", International IC Ð Taipei ¥ Conference Proceedings, Nov,2012